\documentstyle[epsf]{l-aa}
\newcommand{\ra}{{\rm A}}
\newcommand{\rd}{{\rm d}}
\newcommand{\om}{{\omega}}
\newcommand{\Om}{{\Omega}}

\renewcommand{\o}{\over}
\newcommand{\p}{\partial}
\newcommand{\be}{\begin{equation}}
\newcommand{\ee}{\end{equation}}
\newcommand{\bd}{\begin{displaymath}}
\newcommand{\ed}{\end{displaymath}}
\newcommand{\bea}{\begin{eqnarray}}
\newcommand{\eea}{\end{eqnarray}}

\newcommand{\vp}{\varpi}
\newcommand{\si}{\sigma}
\newcommand{\al}{\alpha}
\newcommand{\cs}{c_{\rm s}}
\newcommand{\half}{\textstyle{1\over 2}}
\newcommand{\gapprox}{\;\rlap{\lower 2.5pt
             \hbox{$\sim$}}\raise 1.5pt\hbox{$>$}\;}
\newcommand{\lapprox}{\;\rlap{\lower 2.5pt
             \hbox{$\sim$}}\raise 1.5pt\hbox{$<$}\;}
\newcommand{\bfg}[1]{\setbox0=\hbox{#1}%
  \kern-.025em\copy0\kern-\wd0
  \kern.05em\copy0\kern-\wd0
  \kern-.025em\raise.0433em\box0}
\begin{document}
   \thesaurus{06    % A&A Section 6: Form. struct. and evolut. of stars
              02.09.1 ; %instab
              02.13.2 ; %MHD
              08.13.1; %stars B
              08.18.1 ; %stars rotation
              06.18.2; %sun interior
              }
   \title{Differential rotation and magnetic fields in stellar interiors}
%   \subtitle{}

% 
   \author{H.C. Spruit \inst{1}}

%   \offprints{}

   \institute{ Max-Planck-Institut f\"ur Astrophysik, Postfach 1523, 
               D-85740 Garching bei M\"unchen, Germany}

   \date{Received }

   \maketitle

%   \maintitlerunninghead{}

%   \authorrunninghead{}

\begin{abstract}
The processes contributing to the evolution of an initially weak magnetic field 
in a differentially rotating star are reviewed. These include rotational 
smoothing (akin to convective expulsion) and a list of about 5 instabilities, 
among them magnetorotational instability, buoyancy instability, and pinch-type 
instabilities. The important effects of thermal and magnetic diffusion on these 
instabilities are analyzed in some detail. The first instability to set in is a 
pinch-type instability. It becomes important in modifying the field 
configuration before magnetic buoyancy-driven instabilities set in. The 
evolution of an initially strong field remains a more open question, including 
the old problem whether dynamically stable magnetic equilibria  exist in stars. 


\end{abstract}

\section{Introduction}

A number of different processes cause stars to rotate differentially. Stars with 
convective envelopes experience braking of their rotation by a magnetic stellar 
wind (Schatzman 1962). This torque acts on the convective envelope, and slows it 
down relative to the radiative core. Secondly, in an evolving star the core 
contracts and spins up, while the envelope expands and slows down. If core and 
envelope were to conserve their angular momentum separately, the core of a giant 
would end up rotating some $10^5$ times faster than its envelope. Finally, the 
Eddington-Sweet circulation in a rotating star causes differential rotation by 
transport of angular momentum. In a steady state, such that dissipation and 
driving of the differential rotation by this circulation balance each other, the 
rotation decreases outward by some 30\% between the core and the surface (Zahn 
1992). 

Opposing this differential rotation are friction processes due to hydrodynamical 
instabilities or magnetohydrodynamical processes. The strength of such friction 
determines how much rotation can be left in the end products of stellar 
evolution, white dwarfs and neutron stars. If they are effective enough to 
maintain an approximately uniform rotation in giants, for example, the observed 
rotation of white dwarfs and pulsars is not a leftover of the rotation of their 
progenitors (Spruit \& Phinney 1998, Spruit 1998). Explosion mechanisms of 
supernovae that rely on rotation (Bisnovatyi-Kogan 1970, LeBlanc \& Wilson 1970, 
Meier et al. 1976, Rampp et al. 1998), and explanations of the morphology of 
planetary nebulae and objects like $\eta$ Carinae in terms of stellar rotation 
(Heger \& Langer 1998) also depend crucially on the degree of differential 
rotation that can persist in a stellar interior. The relevant hydrodynamical 
processes  have been classified (Zahn 1974, 1983). Their net effect on angular 
momentum transport is somewhat uncertain and in stellar evolution calculations 
is usually parametrized with arbitrary adjustable coefficients. 

An important test case for such parametrizations is the internal rotation of the 
Sun, which has been measured with remarkable precision though helioseismology 
(Corbard et al. 1997, Schou et al. 1998). The most striking result of these 
measurements is the near-uniform rotation of the radiative core of the Sun. Over 
a depth range of only 0.05$R_\odot$, the differential rotation in the convective 
envelope (30\% faster at the equator than at the poles) changes to a state of 
uniform rotation in the core. Though the measurements are less accurate in the 
inner core ($r/R_\odot<0.3$) and near the rotation axis, no deviation from 
uniform rotation has yet been detected in the core. 

It has been known for a long time that such a low degree of differential 
rotation is incompatible with the currently known hydrodynamic transport 
mechanisms of angular momentum. Circulation is ineffective for a slow rotator 
like the Sun. The known hydrodynamic instabilities are also unlikely to 
contribute, since the conditions for their occurrence are not satisfied in a 
core rotating as slowly and uniformly as the present Sun (Spruit et al. 1983).

It is therefore natural to look for magnetohydrodynamic mechanisms of angular 
momentum transport in stably stratified layers of stars. This may sound like a 
daunting prospect, given the reputation of magnetic fields for complexity and 
lack of firm results. The situation in the context of the present problem is in 
fact quite promising. Over the past 4 decades, a substantial MHD literature 
bearing on the problem has developed. In this paper I first review the various 
contributing processes, conclude that pinch-type instability of an azimuthal 
field (Tayler 1973) is of particular relevance. Then I show that the effects of 
magnetic and thermal diffusion are quite important, and how they can be 
quantified using Acheson's (1978) dispersion relation. 

\subsection{statement of the problem}
Suppose we start with a differentially rotating star with rotation rate $\Om(r)$ 
and field configuration ${\bf B}_0$. We wish to know how the field and the 
rotation evolve in time, and to answer questions like: is the field able to make 
the star rotate uniformly? If so on what time scale? What kind of field 
configuration remains at large time? Though these questions can not be answered 
fully because of a few missing pieces of theory, it turns out that a fairly 
detailed picture can be drawn of the magnetohydrodynamics of an initially weak 
field in a differentially rotating star. In most of the analysis, I will assume 
that there is no external torque on the star, and that its internal structure is 
not evolving. These assumptions are made for convenience and clarity in 
delineating the processes involved; they can be easily relaxed. 

A number of different MHD processes are involved in the problem stated. I 
introduce them by starting with a number of unrealistic simplifying assumptions 
and then relaxing these. The evolution of the field and the rotation depends on 
the strength of the initial field ${\bf B}_0$. In most of the following, I 
assume the initial field to be weak, in the sense that the Alfv\'en travel time 
through the stars is much longer than the rotation period. In terms of the 
Alfv\'en frequency $\Om_\ra$,
\be \Om_\ra = {\bar B\over R(4\pi\bar\rho)^{1/2}},\ee
where $\bar\rho$ is the mean density and $\bar B$ a mean field strength, the 
assumption is that $\Om_\ra\ll\Om$. For the current Sun, this is satisfied if 
the field is small compared to a megaGauss, which in all likelihood is the case. 
For early type stars, with their larger rotation rates, the assumption is even 
more likely to be justified. It is less well justified, however, in the cores of 
giants, if these corotate with their envelopes.

\section{Winding-up of weak fields (no diffusion, $v_\phi$ only)}
\label{nofi}
\subsection{winding-up}
\label{win1}
Start with ignoring magnetic diffusion, so that the induction equation is
\be \p_t{\bf B}=\nabla\times({\bf u}\times{\bf B}),\ee
and by assuming that the only motions to be considered are axisymmetric, purely 
azimuthal motions,
\be v_\phi=r\Om(r,\theta).\ee
Thus we are ignoring for the moment the motions in the $(r,\theta)$ plane due, 
for example, to instabilities. Also ignored are viscosity and thermal diffusion. 
Consider first the case of an {\em axisymmetric} magnetic field (aligned with 
the rotation axis).
In the usual way, an axisymmetric field ${\bf B}$ can be written in terms of a 
stream function $\psi$:
\be 
B_{r}={1\over r^2\sin\theta}\p_\theta\psi,\qquad B_{\theta}=-{1\over 
r\sin\theta}\p_r\psi.
\ee
This function is constant along field lines. Each value of $\psi$ labels a 
different magnetic surface (surface generated by rotating a field line around 
the axis).

The induction equation then has the components
\be 
\p_t B_r=\p_t B_\theta=0,\qquad \p_t B_\phi= r\sin\theta{\bf B}_{\rm 
p}\cdot\nabla\Om.
\ee
Under the assumptions made the poloidal field component ${\bf B}_{\rm 
p}=(B_r,B_\theta,0)$ does not change in time, and the equation for $B_\phi$ can 
be integrated in time:
\be B_\phi=N{\bf n}\cdot {\bf B}_{\rm p}, \ee
where
\be N(r,\theta,t)=r\sin\theta\int_t \vert\nabla\Om\vert \ee
is (modulo a factor $2\pi$), the number of `differential turns', or the 
rotational displacement, and
\be {\bf n}=\nabla\Om/\vert\nabla\Om\vert \ee
is a unit vector in the direction of the gradient of $\Om$.
The torque per unit surface area, 
\be \tau=r\sin\theta B_rB_\phi/4\pi \ee
increases linearly with the displacement $N$, hence the torque causes the 
rotation to execute a harmonic oscillation. Under the assumptions made, the 
oscillation remains linear even though the displacement can be very large. This 
is because of the linear behavior of a pure Alfv\'en wave at arbitrary 
amplitude. The oscillation is, in fact, an Alfv\'en wave traveling along the 
magnetic surface generated by the poloidal field lines with a given value of the 
stream function $\psi$. Its period is the Alfv\'en travel time $\int \rd 
s/V_{\rm Ap}$ along this field line, where $V_{\rm Ap}$ is the Alfv\'en speed 
based on the poloidal field strength (Mestel 1953). 

Considering the evolution of the field on a given time scale $t_0$ (for example 
the age of the star since its formation), we can define a critical field 
strength such that the oscillation period $t_\ra=R/v_\ra$ equals $t$:
\be B_{\rm M}=(4\pi\rho)^{1/2}R/t_0, \ee
where $\rho$ is a typical density in the star, and $R$ its radius.
For example, take for $t_0\approx 10^{17}$s the age of the present Sun, and for
$\rho$ its mean density $\approx 1$. Then $B_{\rm M}\approx 2\mu$G (Mestel 
1953). That is, if the initial field strength is more than a $\mu$Gauss or so, 
the field eventually gets wound up to such an extent that Lorentz forces start 
affecting the internal rotation of the Sun.

\subsection{phase mixing}
If the initial field is large compared with $ B_{\rm M}$, not a very strong 
condition, the rotation on each magnetic surface oscillates around a mean value 
given by the total angular momentum on this surface. Since Alfv\'en waves do not 
couple across magnetic surfaces, neighboring surfaces oscillate independently, 
have different periods, and the oscillations on them increasingly get out of 
phase with each other. The distance between points with oscillation phase 
differing by $\pi$, say, decreases as $1/t$. After a finite time, this distance 
becomes short enough that magnetic diffusion starts becoming important. Thus, it 
is necessary to include magnetic diffusion into the picture.

\section{Winding up and oscillation with magnetic diffusion ($v_\phi$ only)}
\subsection{Rotational smoothing}
\label{smooth}
We now relax the asumption that the initial field is axisymmetric, and also 
include magnetic diffusion. The induction equation is then 
\be \p_t{\bf B}=\nabla\times({\bf u}\times{\bf B})-\eta\nabla^2{\bf B},\ee
where for simplicity I have assumed that the magnetic diffusivity is independent 
of position. This is a good approximation, since the diffusion effects 
encountered here are effective only on small length scales. The poloidal field 
can be decomposed into two components
\be {\bf B_{\rm p}}={\bf B}_{\rm a}+{\bf B}_{\rm n},\ee
where ${\bf B}_{\rm a}$ is the azimuthal average of ${\bf B}$, i.e. $\p_\phi{\bf 
B}_{\rm a}=0$, and ${\bf B}_{\rm n}$ the non-axisymmetric part, whose azimuthal 
average $\langle{\bf B}_{\rm n}\rangle_\phi$ vanishes. Let these components 
initially be of similar strength. 

If the initial field is weak, the field lines of the initial poloidal field $\bf 
B_{\rm p0}$ are wound up tightly before the restoring Lorentz forces become 
effective. Let 
\be q\equiv r\vert\nabla\Om\vert/\Om \label{q}\ee
be a dimensionless measure of the local rate of differential rotation. If the 
initial poloidal field is smooth and varying on a length scale of the order of 
the radius of the star, $q\sim O(1)$. Since the induction equation is linear, 
the winding up process can be considered separately for the components ${\bf 
B}_{\rm a}$ and ${\bf B}_{\rm n}$. Since the azimuthal average of $\bf B_{\rm 
n}$ vanishes, it has opposite polarities as a function of azimuth at any given 
$(r,\theta)$, and the azimuthal angle between opposite polarities is $\pi$ or 
less. Points with field of opposite polarity at the same $\theta$ but separated 
by a small distance $l$ in the direction of the rotation gradient, will be 
brought to the same azimuthal angle after a time $t$ given by:
\be t\Om q l\approx\pi r. \ee
The time scale on which these opposite polarities will cancel each other by 
magnetic diffusion $\eta$ is
\be t_\rd=l^2/\eta. \ee
This cancellation leads to a decay of the poloidal field, governed by the 
approximate equation
\be \p_t\ln B_{\rm n}\approx 1/t_\rd=\eta\left({t\Om q\over \pi r}\right)^2, \ee
which integrates to
\be B_{\rm n}=B_{\rm n0} e^{-(t/t_\Om)^3},\ee 
where the {\it rotational smoothing time} is
\be t_\Om=\left({3r^2\pi^2\over\eta\Om^2q^2}\right)^{1/3}. \label{tom}\ee
Because of the steep dependence on $t$, the non-axisymmetric component of the 
poloidal field effectively disappears after only a few times $t_\Om$.

This process has been studied extensively in the context of the kinematic 
evolution of magnetic fields in convection, and is called convective expulsion 
there (Zeldovich 1956, Parker 1963, Weiss 1966). In the present context of 
differentially rotating stars, it was studied by R\"adler (1980, 1986). The 
remarkable effect of differential rotation on a weak field is thus to {\it expel 
the non-axisymmetric field components} from the star on a finite time scale. For 
example, for an initial Sun rotating at $\Om_0=3\,10^{-5}$ (ten times the 
present-day rate), $q\approx 1$, with a diffusivity $\eta\approx 10^3$, $t_\Om$ 
is only $\approx 100$yr. If nothing else were to happen, the net result of 
differential rotation would be to make the field axisymmetric on a time scale 
which is very short compared to its life time. 

This, however, applies only if the initial field is sufficiently weak, so that 
the winding-up can be treated as a kinematic process. A field is weak in this 
context if magnetic torques, which affect the differential rotation, do not 
become effective during the rotational smoothing process. This is the case if 
the initial Alfv\'en travel time
\be t_{{\rm A}0}=r(4\pi\rho)^{1/2}/B_0 \ee
satisfies
\be t_{{\rm A}0}>t_\Om, \ee
which translates into a condition on the initial field strength:
\be B_0<B_1, \label{cb1}\ee
where
\be B_1=r(4\pi\rho)^{1/2} \left({\eta\Om^2q^2\over 
3r^2\pi^2}\right)^{1/3}.\label{B1}\ee
For the  same initial Sun, $B_1\approx 30$G. In terms of the corresponding 
Alfv\'en frequency:
\be 
{\Om_{{\rm A}1}\o\Om}=\left({q^2\o 3\pi^2}\right)^{1/3} 
\left({\eta\o\Om r^2}\right)^{1/3}.\label{B1a}
\ee
If the poloidal field satisfies (\ref{cb1}), the non-axisymmetric field 
component is smoothed out on the time scale (\ref{tom}), after which there 
remains an axisymmetric field (R\"adler, 1986). For this to be the case, the 
initial Alfv\'en travel time has to be long compared with the rotation period by 
a factor $(\Om r^2/\eta)^{1/3}$. If the initial field is larger than (\ref{B1}), 
rotational smoothing does not happen and instead the differential rotation is 
damped out by the process of phase mixing (all of this still under the 
assumption that $v_\phi$ is the only velocity component). 

\subsection{Phase mixing}
Assume first that condition (\ref{cb1}) is satisfied, so that rotational 
smoothing is effective. How does the field then evolve after being 
axisymmetrized? Since the smoothing process is linear, the axisymmetric and 
non-axisymmetric field components evolve independently. The symmetric component 
continues to be wrapped up after the non-axisymmetric component has been 
smoothed away. Eventually, if time permits, the azimuthal field will become 
strong enough to oppose the wrapping. Phase-mixing then starts acting. 
Oscillations on neighboring magnetic surfaces get out of phase, and the length 
scales in the Alfv\'en wave field become increasingly smaller, until magnetic 
diffusion becomes effective and damps out the wave [Spruit 1987, Roxburgh 1987 
(private communication)]. If the Alfv\'en travel time varies through the star by 
a factor $q_\ra=\Delta t_\ra/t_\ra$, the phase difference between magnetic 
surfaces separated by a distance $l$ is of the order $ \Om_{\rm A} t q_\ra l/r$. 
After a time $t$ this phase difference reaches a value of $\pi$ if the length 
scale $l$ is
\be {l\over r}={\pi \over q_\ra\Om_\ra t}.\ee
The magnetic diffusion time on this length scale is $t_\rd=\l^2/\eta$. The wave 
amplitude $A$ gets damped by magnetic diffusion on this time scale,
\be \p_t\ln A=1/t_\rd=\eta \left({t\Om_\ra q_\ra\over \pi r}\right)^2.\ee

This is the same equation as in the case of rotational smoothing, but with the 
wave frequency $\Om_\ra$ replacing the rotation rate $\Om$. The wave amplitude 
decays like
\be A=A_0e^{-(t/t_{\rm p})^3},\ee
where the {\it phase mixing time scale} $t_{\rm p}$ is
\be t_{\rm 
p}=\left({3\pi^2r^2\over\eta\Om_\ra^2q_\ra^2}\right)^{1/3}.\label{omp} \ee
When the oscillations have damped out, the rotation rate is constant on each 
magnetic surface:
\be \Om=f(\psi), \qquad (t\gapprox t_{\rm p}).\ee
The phase mixing process in Alfv\'en waves also plays a role in other 
astrophysical situations, see Heyvaerts \& Priest (1983), Sakurai \& Granik 
(1984), Petkaki et al. (1998). It is analogous to the phase mixing process in 
plasma physics only in a rather broad sense.

Expression (\ref{omp}) for the time scale is very similar to that for rotational 
smoothing (\ref{tom}), but its consequences are rather different. Rotational 
smoothing removes the non-axisymmetric components of the magnetic field, phase 
mixing damps out the variations of $\Om$ along magnetic surfaces. This has been 
studied with numerical examples by Mestel, Moss \& Tayler (1988, 1990), and Moss 
(1992). Detailed calculations of the phase mixing process have been made by 
Charboneau \& McGregor (1993) and Sakurai et al. (1995). 

\subsection{Stronger initial fields, and interim conclusions}
If $B_0>B_1$ (Eq.\ \ref{B1}), torques become effective before rotational 
smoothing can make the field axisymmetric. If the initial field is 
non-axisymmetric, it will therefore stay non-axisymmetric. Such a field still 
has magnetic surfaces however. Alfv\'en waves and the slow mode continuum travel 
on these surfaces (Goossens et al. 1985), causing phase mixing just as in the 
axisymmetric case. These waves will damp out on the time scale $t_{\rm p}$. The 
final field will then be similar to the initial field. But what is the state of 
rotation at the end of the process? Since the magnetic surfaces are not 
axisymmetric, and differential motions are damped on these surfaces, the 
stationary state is one of {\it uniform rotation}, and nonuniform rotation is 
possible only if the initial field is axisymmetric (Mestel et al. 1988, 1990). 

This completes the discussion for the case when only the azimuthal equation of 
motion is taken into account. Depending on the initial field strength, the final 
state in this picture is either a uniformly rotating magnetic star with a 
non-axisymmetric field, or a differentially rotating star with an axisymmetric 
poloidal field, with rotation constant on magnetic surfaces. This is only a 
preliminary scenario, of course, since we have allowed only for purely azimuthal 
motions. The axisymmetric poloidal end-state, if it ever were to materialize, 
would evolve further by magnetic instabilities. 

\section{Magnetic instabilities}
The fluid motions, taken to be purely azimuthal in the above, are now 
unconstrained, and the evolution of the field becomes much more interesting 
since magnetic instabilities can take place. 

Consider first the case where the initial field is weak, such that differential 
rotation has time to produce a predominantly azimuthal field by rotational 
smoothing (Sect.\ \ref{smooth}). Stronger initial fields are taken up again in 
Sect.\ \ref{strong}. The restriction to weak initial fields isolates the 
magnetic instabilities that appear from the phase mixing process discussed 
above, which can be ignored here.

\subsection{Magnetic shear instability}
\label{mshear}
In a differentially rotating object with angular velocity decreasing outward, a 
weak field is generically unstable to magnetorotational instability (Velikhov 
1959, Chandrasekhar 1960, Fricke 1969, Acheson 1978, Balbus \& Hawley 1991, 
1992, Balbus 1995). This is a linear instability in which the magnetic field 
mediates the release of free energy in differential rotation. The growing 
magnetic perturbations couple the fluid at different radii, the differential 
rotation acting on them converts the free energy into magnetic and kinetic 
energy on smaller scales. The instability acts when the rotation rate decreases, 
but angular momentum increases outward. In other cases, i.e. when angular 
momentum ($J$) decreases, or angular velocity increases outward, ordinary 
hydrodynamic shear instabilities are known to exist. These cases are less 
relevant for differentially rotating stars or accretion disks, however. Whether 
purely {\it hydrodynamic} instabilities exist for $\Om$ decreasing and $J$ 
increasing with radius is still controversial. The possibility that this case 
might actually be hydrodynamically {\it stable} should be taken seriously, 
however, given that instabilities have so far been found neither theoretically, 
nor numerically, nor experimentally. A physical argument for stability, not 
constituting a proof, has been given by Balbus et al. (1996). 

Where magnetic shear instability is present, it generates a magnetic form of 
turbulence. In stars, a strongly stabilizing factor is the stratification. 
Except in regions close to convective instability, this limits the instability 
to cases with strong differential rotation, and a rotation rate near the maximum 
value. Apart from these extreme situations, magnetic shear instability is 
limited to displacements on horizontal surfaces, and redistributes angular 
momentum over such a surface if the rotation rate decreases with cylindrical 
radius (Kato 1992, Balbus 1995). The effect of the instability is thus much more 
limited than in the case of an accretion disk, where differential rotation is 
dominant, pressure effects small, and magnetic shear instability endemic. 

On sufficiently small length scales, however, thermal diffusion can strongly 
reduce the stabilizing buoyancy effect of the thermal stratification. As in the 
case of purely hydrodynamic instabilities (Zahn 1974, 1983), one would therefore 
expect that {\it i)} linear magnetic shear instability should reappear at large 
wavenumbers when thermal diffusion is taken into account, and {\it ii)} angular 
momentum transport by some form of small scale magnetic turbulence will take 
place. The first of these was already demonstrated to be the case in the 
classical and detailed analysis by Acheson (1978). He showed that the necessary 
and sufficient condition for linear instability is 
\be q=-\rd\ln\Om/\rd\ln r>{N^2\over 2\Om^2}{\eta\over\kappa},\label{ach}\ee
(cf. eq \ref{q}), where $N$ is the buoyancy frequency and $\kappa$ the thermal 
diffusivity. For conditions in stellar interiors, $\kappa\gg\eta$. This is 
because the magnetic diffusion involves the random walk of electrons, which have 
much shorter mean free paths than the photons that carry the heat. Thus magnetic 
shear instability is possible with weak rotation gradients 
$\vert\rd\ln\Om/\rd\ln r\vert\ll 1$, provided that the rotation rate is not too 
slow compared with the buoyancy frequency. For a star rotating as slowly as the 
Sun, however, a fairly strong gradient in rotation rate is still needed for 
instability. For the present Sun, with $N\sim 10^{-3}$, $\Om=3\,10^{-6}$, 
$\eta/\kappa\sim 5\,10^{-5}$, instability would require $q\sim O(1)$. This is a 
much larger degree of differential rotation than allowed by current observations 
(Schou et al. 1998) of the Sun's interior rotation.

How much angular momentum transport this process is likely to produce, in cases 
where a large enough shear exist,  is not certain at the moment. The issue is 
discussed in Spruit \& Balbus (in prep.). Several lines of argument presented 
there suggest that a modestly effective transport of angular momentum is 
possible. A simple, perhaps convincing, argument starts by assuming that the 
magnetic turbulence has the effect of enhancing the magnetic diffusion of large 
scale fields. The amplitude at which this turbulence saturates is then found by 
requiring that instability is just possible according to condition (\ref{ach}), 
with the magnetic diffusivity $\eta$ replaced by an effective diffusivity 
$\eta_{\rm e}$. This condition then determines $\eta_{\rm e}$:
\be \eta_{\rm e}=2q\Om^2\kappa/N^2.\label{etef}\ee
[Recall that the same argument, applied to convective instability, can be used 
to derive the familiar mixing-length model of convection.] The fluid motions in 
the turbulence are of slow-mode and Alfv\'enic type, so that the kinetic energy 
is of the same order as the energy in the magnetic field. If $\nu_{\rm e}$ is 
the effective viscosity due to the fluid motions, the magnetic Prandtl number 
$P_{\rm e}=\nu_{\rm e}/\eta_{\rm e}$ is then of order unity. Numerical 
simulations suggest $P_{\rm e}\approx 0.1$ (Hawley et al. 1995, Brandenburg et 
al. 1995, Matsumoto \& Tajima 1995). With the assumptions made, (\ref{ach}) now 
yields the effective viscosity expected:
\be \nu_{\rm e}=2P_{\rm e}q\kappa{\Om^2\over N^2}. \label{loop} \ee

Since the induction equation is linear, the winding-up, rotational smoothing and 
phase mixing processes discussed above all just proceed as before, the only 
difference being that the magnetic diffusion has to be replaced by the effective 
value (\ref{etef}). Quantitatively, the difference is not very large. For the 
initial Sun assumed above, the effective diffusivity in the presence of magnetic 
shear instability would have been only 10--100 times larger than the microscopic 
value. With this effective diffusivity, the rotational smoothing time 
(\ref{tom}) decreases only by a factor of order unity. 

The net effect of magnetic shear instability therefore may be small. Even though 
it is not always suppressed by a stable thermal stratification, its effects are 
modest. This conclusion, however, is preliminary, as it depends on an estimate 
of its nonlinear development. Numerical simulations should be able to settle 
this question more definitively.  

\subsection{Instabilities of an azimuthal magnetic field}
During the rotational smoothing and phase mixing processes in an initially weak 
field, the azimuthal field is much larger than the poloidal component. An 
azimuthal field has its own instabilities. Magnetic shear instability operates 
also on such an azimuthal field, but is more properly regarded as a form of 
shear instability, extracting its energy from differential rotation (as 
discussed above). Other forms of instability exist, which draw their energy from 
the magnetic field itself. They are of two distinct types. Perhaps the most 
well-known is magnetic buoyancy or Parker instability (Parker 1966), due to a 
vertical gradient in the field strength. This is discussed in Sect.\ \ref{buoy}. 
The second is a pinch-type instability.

\section{Tayler instability}
\label{horg}
The second form of instability derives its energy from (nearly) horizontal 
interchanges. The classical result in this context is that of Tayler (1973), who 
considered adiabatic perturbations ($\eta=\kappa=0$) in a stratified, 
nonrotating star. By an energy method he showed that {\it every purely azimuthal 
field} $B_\phi(r,\theta)$ in a stably stratified (nonrotating) star is {\it 
unstable on an Alfv\'en time scale}, no matter how weak the magnetic field. This 
form of instability is closely related to many forms of instability in pinches 
and torus configurations (e.g. Tayler 1957). Its behavior under the strongly 
stratified conditions in stellar interiors is sufficiently distinct, however, 
that I find it convenient to refer to it by the separate name of `Tayler 
instability'. In poloidal field configurations, pinch-type instabilities also 
occur in stellar interiors; they were studied by the same method by Wright 
(1973), see also Sect.\ \ref{pol}.

\subsection{Previous results}
The most unstable motions are $m=0$ and $m=1$ modes with nearly horizontal 
displacements. The displacements take place at nearly constant total pressure, 
i.e. the Eulerian perturbation $\delta (P+B^2/8\pi)$ vanishes. They are 
essentially local in the $r$ and $\theta$ directions: when the instability 
condition is satisfied at a point $(r,\theta)$, unstable perturbations can be 
found that are confined to a small neighborhood of this point. In spherical 
coordinates (Goossens et al. 1981; Tayler's was derived in cylindrical 
coordinates), the necessary and sufficient conditions for instability are
\be 
\p_\theta \ln (B^2\sin\theta\cos\theta)>0, \qquad (m=1)\label{m=1}
\ee
\be \cos\theta\,\p_\theta \ln (B^2/\sin^2\theta)>0, \qquad (m=0)\label{m=0}\ee
From the induction equation it is evident that azimuthal fields created by 
winding up of a poloidal field must have
\be B\sim\theta\qquad (\theta\ll 1)\ee
near the pole. Hence there is always a region near the pole where  condition 
(\ref{m=1}) is satisfied, and the field is unstable. Away from the pole, the 
most unstable modes can be either $m=1$ or $m=0$. As an example, consider the 
field resulting from the winding up of an initially uniform field parallel to 
the rotation axis, when rotation is a function of $r$ only. This field has
\be B_\phi\sim \sin\theta\cos\theta.\ee 
The region $\pi/4<\theta<3\pi/4$ is stable for $m=1$ modes, the polar caps are 
unstable, and the axisymmetric modes are stable. 
The growth rate is of the order (Tayler 1973, Goossens et al. 1981)
\be \sigma\approx \Om_\ra=V_\ra/r\qquad(\Om\ll\Om_\ra).\ee

\begin{figure}
\mbox{\epsfxsize 8cm\epsfbox{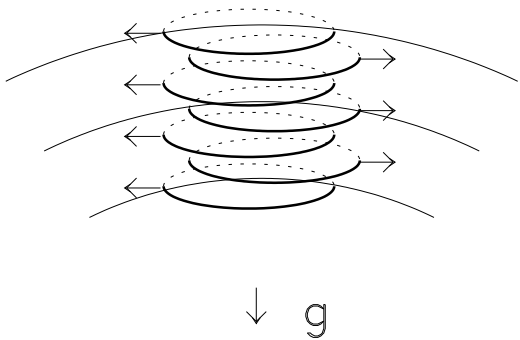}}\hfil\mbox{}
\caption{\label{displ}Unstable displacements in an azimuthal field near the 
pole. Shown is the $m=1$ mode, which occurs under the widest range of 
conditions. The displacements are along horizontal surfaces (indicated by arcs)}
\end{figure}

The unstable displacements are sketched in Fig.~\ref{displ}.  

When rotation is included, the energy method does not work, and somewhat weaker 
results are obtained. Nevertheless, Pitts \& Tayler (1986) showed that rotation 
does not by itself remove the instability. There are still unstable $m=1$ modes, 
though their growth rate is reduced. When rotation is rapid compared with the 
Alfv\'en travel time, as I have assumed above, the growth rate found by Pitts 
and Tayler is of the order 
\be \sigma\approx\Om_\ra^2/\Om\qquad (\Om\gg\Om_\ra).\label{pt}\ee 
This dependence on the rotation rate is typical for instabilities of an 
interchange type. It represents the fact that the Coriolis force, being 
perpendicular to the velocity, does not enter into the energy budget of the 
perturbations, though it can still affect the shape of the unstable modes and 
their growth rates.

The growth rate (\ref{pt}) is still fast compared with the rotational smoothing 
and phase mixing time scales, so that the effects of the instability are likely 
to significantly change the evolution of an initially weak field as sketched in 
Sect.\ \ref{nofi}.

\subsection{Effects of rotation and diffusion}
\label{tay}
The published results deal mostly with the stability and growth rates 
considering the star as a whole, using an energy method. Tayler's result shows, 
however, that the essence of the instability is a local interchange-type process 
(see Appendix 2 in Tayler 1973). Hence it is natural to study the instability by 
a modal analysis, in a local approximation.  This also has the advantage that 
the effects of rotation, magnetic and thermal diffusion can be included, which 
is not possible with an energy method. 

The result of such a local approximation, for a purely azimuthal field, has been 
 given by Acheson (1978). His dispersion relation includes the effects of 
rotation, viscosity, thermal and magnetic diffusion, and is rather complicated. 
The special case of an azimuthal field at the equator has been analysed in 
detail by Acheson (1978, 1979). Tayler instability, however, disappears at the 
equator, and shows its most characteristic behavior at the poles. 

In the following I derive the properties of the instability by heuristic 
arguments. In the Appendix, I derive these results more systematically from the 
dispersion relation. I also show there that Acheson's local approximation 
agrees, in cases where a comparison is possible, with all results based on the 
more rigorous energy method. This establishes the validity of his local analysis 
for the case of Tayler instability. The Appendix also discusses a few subtleties 
that are not captured by the heuristic derivations.

A useful simplification is to ignore viscosity. In a stellar interior in which 
radiation is not the dominant source of viscosity, it is smaller than the next 
larger diffusion process, magnetic diffusion, by 1--2 orders of 
magnitude\footnote{The effects of viscosity may need to be reconsidered, 
however, for the interiors of massive stars, where radiation pressure is 
important.}. Use cylindrical coordinates $(\vp,\phi,z)$ at the pole, where the 
rotation axis is in the $z$ direction. The magnetic field is ${\bf 
B}=(0,B(\vp),0)$.  Define an Alfv\'en frequency $\om_\ra$:
\be \om_\ra={B\over(4\pi\rho)^{1/2}\vp}, \ee
(which is a function of $\vp$, as opposed to the very similar quantity $\Om_\ra$ 
used above). Since the analysis is local, the field is characterized completely 
by two numbers, the local value $B$ of the field strength, and its radial 
gradient
\be p\equiv \rd\ln B/\rd\ln \vp. \ee
The vertical gradient of the field is neglected here. Instabilities associated 
with such a gradient are discussed below in Sect.\ \ref{buoy}. The rotation rate 
is assumed to be uniform in the present analysis. A gradient $\rd\Om/\rd z$ is 
present and causes its own forms of instability, but these have been covered 
already above in the discussion of magnetic shear instability.

There are several different time scales involved in the problem. The largest 
frequency is the buoyancy frequency $N$, reflecting the strong effect of 
stratification. The next lower frequency is the rotation rate $\Om$. For the 
weak fields that turn out to be most relevant, the Alfv\'en frequency $\om_\ra$ 
is small compared to $\Om$. In this case we have the ordering
\be N\gg\Om\gg\om_\ra.\ee

The perturbations are of the form 
\be e^{i(lr+m\phi+nz)+\sigma t},\ee
and are essentially incompressive\footnote{Small expansions and contractions 
occur in order to make the total pressure perturbation vanish for the most 
unstable modes. These disappear in the limit $v_\ra/c_{\rm s}\approx 
\om_\ra/N\ll 1$.}, $\rm{div}{\bf v}=0$. In the absence of diffusion 
($\kappa=\eta=0$), the growth rate of the instability is maximized for 
$\/n\rightarrow 0$ (Pitts \& Tayler 1986). In this limit, the instability is 
essentially confined to horizontal surfaces. The growth rate, which in the 
absence of rotation is of the order $\om_\ra$, is reduced by the Coriolis forces 
to a value of the order 
\be \sigma\sim \om_\ra^2/\Om, \qquad (\Om\gg\om_\ra)\label{oma}\label{adiab}\ee
but the instability condition itself does not depend on $\Om$. This behavior is 
generic for interchange-type instabilities in the presence of rotation, as 
discussed above. The instability conditions are (Tayler 1957, see Appendix)
\be p>{m^2\o 2}-1\quad (m\ne 0), \qquad {\rm and}\qquad p>1\quad (m=0). \ee
The $m=1$ mode thus occurs under the widest range of conditions (usually, its 
growth rate is also largest if several modes are unstable). In the following, I 
restrict the discussion to $m=1$. 

\subsubsection{Magnetic diffusion only}
\label{wma}

Consider first the case where magnetic diffusion is included but thermal 
diffusion is neglected. This would be appropriate for situations where the 
dominant contribution to the buoyancy frequency is a composition gradient.  In 
order to avoid doing work against the stable stratification, the unstable 
displacements must be nearly horizontal, $v_z/v_{\rm p}\sim l/n\ll 1$. For 
displacements of amplitude $\xi$, the work done per unit mass against the stable 
stratification is $\half\xi^2(l/n)^2N^2$. The energy gained from the field 
configuration is $\half\om_\ra^2\xi^2$. For instability, the field must be 
strong enough, such that $\om_\ra^2>({l^2/ n^2})N^2$. For a given vertical 
wavenumber, this is most easily satisfied at the longest possible horizontal 
wavelength, which is of the order $r$, the (spherical) radius. The vertical 
wavenumber thus has to satisfy
\be n^2>{N^2\over\om_\ra^2r^2}.\label{c1}\ee
At large wavenumbers, however, diffusion starts affecting the perturbations. The 
condition that the rate at which they decay by magnetic diffusion does not 
exceed the growth rate yields
\be n^2\eta<\sigma,\label{c2}\ee
where $\sigma$ is the growth rate in the absence of stratification and 
diffusion. For our ordering $\Om\gg\om_\ra$, $\sigma$ is given by (\ref{oma}). 
Wavenumbers for which both conditions (\ref{c1},\ref{c2}) are satisfied then 
exist if 2 conditions are satisfied. The first is
\be p>-{1\o 2},\ee
which is always satisfied in some region near the poles. The second is
\be 
\om_\ra>\Om({N\over\Om})^{1/2} ({\eta\over r^2\Om})^{1/4}.
\qquad (\Om\gg\om_\ra,~\kappa=0)\label{dif1}
\ee
This condition applies to the most unstable mode, $m=1$ . For example, in the 
outer core of the present Sun ($r\sim 5\,10^{10}$, $N\sim 10^{-3}$, $\eta\sim 
2\,10^3$, $\kappa\sim 4\,10^7$, $\Om=3\,10^{-6}$), the minimum azimuthal field 
strength needed for instability, $B=\om_\ra r(4\pi\rho)^{1/2}$, would be of the 
order $1000$G. 

\subsubsection{Magnetic and thermal diffusion}
Condition (\ref{dif1}) overestimates the field strength required for instability 
if the stratification is due to a thermal gradient. At the short wavelengths 
where magnetic diffusion plays a role, thermal diffusion strongly reduces the 
stabilizing temperature perturbations. As in the case of hydrodynamic 
instabilities in a stable thermal stratification (Zahn 1974), this effect can be 
taken into account by replacing $N$ with an effective buoyancy frequency $\tilde 
N$,
\be \tilde N^2=N^2/(1+\tau/\tau_{\rm T}), \label{zahn}\ee
where $\tau_{\rm T}=(n^2\kappa)^{-1}$ is the thermal diffusion time at 
wavenumber $n$, and $\tau$ the time scale of the process under consideration. In 
our case $\tau$ is the adiabatic instability time scale $\sigma^{-1}$ (Eq.\ 
\ref{adiab}). For these time scales, the temperature perturbations are reduced 
by the factor in brackets in (\ref{zahn}). Since $\kappa\gg\eta$, this factor is 
approximately $\tau/\tau_{\rm T}$. The same argument as that leading to 
(\ref{dif1}) then yields as conditions for instability (cf.\ Appendix eq 
\ref{adif2}) $p>-{1\o 2}$ and
\be 
{\om_\ra\over\Om}>\left({N\o\Om}\right)^{1/2} \left({\eta\o\kappa}\right)^{1/4} 
\left({\eta\o r^2 N}\right)^{1/4}. ~(\Om\gg\om_\ra,~ \kappa\gg\eta)\label{dif2} 
\ee

With this condition the critical field for instability in the present Sun is now 
about 100G, and the typical growth time (from \ref{oma}) is of the order $10^5$ 
yr (for conditions significantly exceeding marginal). The wavenumber at marginal 
stability is $nr=10^3$, corresponding to a wavelength of 3000km. 

\subsubsection{Effect of a composition gradient}
The situation is more complicated if the stratification is due to the combined 
effects of a thermal gradient and a gradient in composition. If $\mu$ is the 
mean atomic weight per particle, the buoyancy frequency is given by (e.g. 
Kippenhahn \& Weigert, 1990):
\be N^2=N_{\rm T}^2+N_\mu^2, \ee
where the thermal and compositional contributions $N_{\rm T}$ and $M_\mu$ are 
given by
\be 
N_{\rm T}^2={g\delta\o H}(\nabla-\nabla_{\rm a}); 
\qquad N_\mu^2={g\phi\o H}\nabla_\mu\label{N}
\ee
with
\be 
\nabla={\rd\ln T\o\rd\ln P}, \quad\nabla_{\rm a}=\left({\p \ln T\o\p\ln 
P}\right)_{S,\mu},\quad \nabla_\mu={\rd\ln\mu\o\rd\ln P}.
\ee
\be 
\delta=\left({\p\ln\rho\o\p\ln T}\right)_{P,\mu},\qquad
\phi=\left({\p\ln\rho\o\p\ln\mu}\right)_{P,T}.
\ee
Here straight derivatives measure the variation of the physical variables with 
depth in the stratification, partials are thermodynamic derivatives of the 
equation of state, and $S$ is the entropy. For an ideal gas, ($P=\rho{\cal 
R}T/\mu$), $\phi=\delta=1$, hence
\be 
N_{\rm T}^2={g\o H}(\nabla-\nabla_{\rm a}); 
\qquad N_\mu^2=g{\rd\ln\mu\o\rd z}.
\ee
For adiabatic perturbations, the buoyancy frequency in the formulas above is 
just replaced by (\ref{N}). For nonadiabatic perturbations, we need to take into 
account that inhomogeneities in composition and temperature diffuse at different 
rates. The  diffusivity $\kappa_\mu$ of inhomogeneities in $\mu$ is of the same 
order as the viscosity, since both quantities scale with the mean free path and 
velocity of the ions. We can then neglect $\kappa_\mu$ compared with the 
magnetic diffusivity. The same arguments as those leading to eq (\ref{dif1}) and 
(\ref{dif2}) then gives the instability conditions $p>-{1\over 2}$ and
\be
{\om_\ra\over\Om}>\left({N_{\rm T}^2\o\Om^2}{\eta\o\kappa} + 
{N_\mu^2\o\Om^2}\right)^{1/4} \left({\eta\o r^2\Om}\right)^{1/4}. \label{dif3} 
\ee
\bd
(\Om\gg\om_\ra,~ \kappa\gg\eta)
\ed

\subsection{Magnetic buoyancy instability}
\label{buoy}
If the magnetic field strength increases in the direction of gravity, the gas is 
supported against gravity in part by the magnetic pressure. If the gradient is 
sufficiently strong, the free energy in the field gradient can be released by 
buckling of the field lines. As with all instabilities where displacements in 
the vertical are necessary, the stratification provides a strong stabilizing 
force against such buckling. Like in the other instabilities, however, this 
stabilizing force is reduced by thermal diffusion on sufficiently small scales. 
To demonstrate the properties of the instability it is sufficient to consider 
the situation at the equator. This case has been analysed in detail by Acheson 
(1978). Ignoring viscosity, the condition for instability is (Acheson's Eq. 
7.27):
\be 
-{\Om^2\over\om_\ra^2}{\rd\ln\Om\over\rd\ln r}-\left({r\over H}-2\right) {\rd\ln 
B\over\rd\ln r}>{\eta\over\kappa}{\gamma N^2\o\om_\ra^2},\label{727}
\ee
where $\om_\ra=V_\ra/r$, $H$ the pressure scale height, and $\gamma$ the ratio 
of specific heats. The first term describes magnetic shear instability, 
discussed above in Sect.\ \ref{mshear}. By assuming uniform rotation, we get:
\be
-\left({r\over H}-2\right) p>{\eta\over\kappa}{\gamma N^2\over\om_\ra^2},
\ee
where $p=\rd\ln B/\rd\ln r$. This condition is independent of the rotation rate. 
The growth rates, however, are reduced by rapid rotation. Compared with the 
nonrotating case, they are smaller by a factor $\sigma/\Om$, where $\sigma$ is 
the growth rate in the absence of rotation. 

Near the center of the star, $r\rightarrow 0$ and $H\rightarrow\infty$, so that 
the second term in the bracket dominates. Instability is then possible if the 
field strength {\it increases} outward. This is obviously not the normal 
buoyancy instability. It is, in fact, the same pinch-type instability as the 
instability that appears near the poles, and is associated there with the 
horizontal gradient of the field strength. These have been discussed in Sect.\ 
\ref{horg}. The directions of the rotation axis and stratification are different 
in the present case, but in the absence of rotation and gravity the instability 
would be the same. 

Hence we can concentrate on here the case $H\ll r$. In terms of the Alfv\'en 
frequency, the instability condition can be written as
\be \om_\ra^2>{\gamma\over -p}{\eta\over\kappa}{H\over r}N^2.\label{buo} \ee
For a smooth field gradient, $-p\sim O(1)$, the critical Alfv\'en frequency for 
instability is of the order (subscript $_{\rm b}$ for buoyancy):
\be \om_{\rm Ab}\approx({\eta H\over\kappa r})^{1/2}N. \ee
Comparing this with condition (\ref{dif2}) for diffusive Tayler instability, 
with its critical Alfv\'en frequency $\om_{\rm AT}$, we have
\be 
{\om_{\rm Ab}\over\om_{\rm AT}}\approx({N\over\Om})^{1/2} 
({\eta\over\kappa})^{1/4}  ({H\over r})^{1/2} ({\Om r^2\over \eta})^{1/4}. 
\label{rat}
\ee
This is minimized for the largest possible rotation rate, $\Om\sim N$. The 
dominant factor is the last one, which is quite large since the magnetic 
diffusivity is so small for stellar length and time scales. One finds that 
$\om_{\rm Ab}/\om_{\rm AT}\gg 1$ for main sequence stars, white dwarfs and 
neutron stars. For the present Sun, $\om_{\rm Ab}/\om_{\rm AT}\approx 10^3$. 

Buoyancy instability thus requires much stronger fields than Tayler instability. 
A field wound up by differential rotation into an azimuthal field therefore 
becomes unstable to Tayler instability first. If the instability is able to 
limit the growth of the toroidal field, a slowly wound-up field will settle at a 
value near the marginal conditions for instability (Mestel \& Weiss 1987), and 
it is unlikely that buoyancy instability will become important. 

\begin{figure}
\mbox{\epsfxsize 8cm\epsfbox{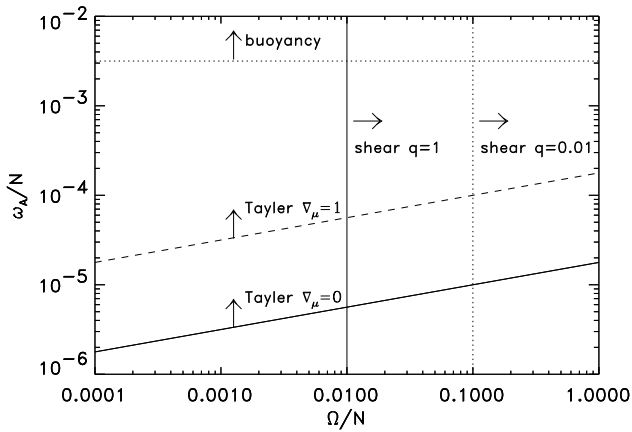}}\hfil\mbox{}
\caption{\label{instb}Stability conditions for an azimuthal magnetic field as 
functions of field strength and rotation rate.  $\om_\ra$ is the Alfv\'en 
frequency $V_\ra/ r$, $N$ the buoyancy frequency, $q$ is the differential 
rotation $\rd\ln\Om\rd\ln r$. Assumed stellar parameters are 
$\eta/\kappa=10^{-4}$, $r^2N/\eta=10^{15}$. With increasing field strength, the 
first instability to appear is Tayler instability. Magnetic shear instability 
appears above a minimum rotation rate.}
\end{figure}

The stability conditions discussed are summarized in Fig.~\ref{instb}

\section{Poloidal field instabilities}
\label{pol}
Though azimuthal fields are conceptually attractive as a natural result of 
differential rotation, the possibility of purely poloidal fields ($B_\phi=0$) or 
general mixed poloidal-toroidal fields must also be considered as possible 
initial field configurations. As discussed further in Sect.\ \ref{strong}, few 
results exist on the stability of mixed poloidal-toroidal fields. 

For purely poloidal fields, strong results are again available. The most 
important result is that, most likely, all purely poloidal fields in stars are 
unstable to adiabatic perturbations, in the absence of rotation. This was 
demonstrated by Wright (1973) and by Markey \& Tayler (1973, 1974). These 
authors considered poloidal fields in which some or all field lines in each 
meridional plane are closed within the star. Then on each of these planes there 
is (at least one) point where the field strength vanishes. Near this point the 
field lines are ellipses centered around the point. The configuration thus 
closely resembles the configuration of an azimuthal field near the pole, and one 
expects similar pinch-type instabilities, driven by the curvature of the field 
lines\footnote{Note that this is now the third instance of such instabilities. 
The first time was instability at the poles, the second time in the instability 
of an azimuthal field near the center of a star in Sect.\ \ref{buoy}.}. The 
situation differs in the present case in that the direction of the stable 
stratification is now within the plane of the field lines. Unstable 
displacements ${\bfg{$\xi$}}$ must be nearly incompressive and close to 
horizontal surfaces, ${\bf g} \cdot\bfg{$\xi$}=0$. Displacements in latitude, 
with a small length scale in the azimuthal direction satisfy these requirements. 
They have the same effect as an $m=1$ displacement in an azimuthal field near 
the pole. Wright (1973) and Markey \& Tayer (1973, 1974) find that these are 
indeed the most unstable ones, and that they  make all poloidal fields with 
closed field lines unstable. The growth rate, as expected, is of the order 
$\om_\ra$.

A special case occurs when none of the field lines is closed inside the star. 
All field lines then cross the stellar surface. An example would be a uniform 
field inside the star, with a dipolar vacuum field outside. The results by 
Markey and Tayler do not apply to this case, but a simple argument (Flowers \& 
Ruderman 1977) shows that this case is equally unstable. As a trial function for 
the displacements consider splitting the star in half by a plane containing the 
axis, and rotating one of the halves over 180$^\circ$. This changes neither the 
thermal, nor the gravitational, nor the magnetic energy of the star. It changes 
the external vacuum field, however. By a suitable choice of the plane, the 
rotation brings fields of opposite polarity closer together on the surface, 
which lowers the energy of the external vacuum field. This is most easily 
visualized for a dipole field. The star is then analogous to a bundle of bar 
magnets, oriented in parallel. By splitting the bundle in two and rotating one 
half, the energy of the bar magnets in the field of the other half is reduced by 
an amount of the order $B^2 V/8\pi$, where $V$ is the volume of the star. Note 
that this energy change is due entirely to the external field energy, the 
internal field energy does not change. With this value for the change in energy, 
the growth rate of the instability is of the order $\om_\ra$. The process can be 
continued by rotations along other planes, which increases the complexity of the 
external field (higher multipoles), and reduces its strength. Asymptotically, 
the external field energy and the growth rate of the instability vanish.

\subsection{Effect of rotation}
For adiabatic perturbations, the effect of rotation has been studied by Pitts 
\& Tayler (1986). The results are less complete than in the nonrotating case, 
since the powerful energy method fails for rotating systems. For the cases 
studied, however, poloidal fields were again found to be unstable. The reason 
for this can be visualized easily by considering the most unstable displacements 
of the nonrotating case. These have azimuthal wavenumber $m\gg 1$ and 
$\xi_\theta\gg \xi_r,\xi_\phi$. The Coriolis force on such displacements is in 
the azimuthal direction. On neighboring meridional planes, such that the phase 
of the perturbation differs by $\pi$, the Coriolis forces are opposite. These 
forces can be balanced entirely by the azimuthal pressure perturbation, so that 
the net effect of the Coriolis force vanishes for high-$m$ perturbations. The 
instability then proceeds under the same conditions and with the same growth 
rate as in the nonrotating case. An explicit example of this effect has been 
given in Spruit \& Taam(1990), where poloidal field instability was studied in 
the very analogous case of a uniformly rotating disk. Numerical simulations of 
this instability in rotating disks have been made by Stehle (1998).

One should therefore expect rotation to have even less effect on poloidal fields 
than it has on the azimuthal fields discussed in Sect.\ \ref{horg}.

\subsection{Effect of magnetic and thermal diffusion}
The effect of the diffusivities has not been studied for poloidal fields. Given 
the nature of the unstable diplacements, which are closely analogous to the 
$m=1$ instabilities near the pole in an azimuthal field, we should expect the 
effects of the diffusivities to be qualitatively the same. In the nonrotating 
azimuthal field case, diffusion does not stabilize the configuration at any 
field strength, but only affects the growth rates. Since the effect of rotation 
on the unstable displacements is small for poloidal fields (see above), it is a 
good guess that poloidal fields will be unstable at all field strengths, even in 
the presence of rotation and magnetic diffusion. Demonstration of this may need 
more detailed study, however.

\section{Initial field strong: stable equilibria?}
\label{strong}
In the above, I have assumed the initial field to be weak, so that a 
predominantly azimuthal field quickly develops by differential rotation. If the 
initial field is strong (as measured by condition \ref{B1}), this is not the 
case. Instead, in this case, the differential rotation can be regarded as a 
perturbation of the field which damps out on the phase mixing time scale 
(\ref{omp}). In doing so, the field settles to a stable equilibrium state, if 
one exists. 

This is a loose end in the story since in spite of extensive work (mainly in the 
60's and 70's, see e.g.  Tayler 1980, Borra et al. 1982, Mestel 1984), it is 
still not known how to prove the existence or absence of stable magnetic 
equilibria in stars, whether they rotate or not. As discussed above, all purely 
toroidal field configurations are {\it unstable} above a critical field strength 
given by the diffusive Tayler instability condition (\ref{dif2}). All purely 
poloidal fields are likely to be unstable as well, as discussed in Sect.\ 
\ref{pol}. For mixed poloidal-toroidal fields, stability analyses exist as well 
(e.g. Wright, 1973, Tayler 1980), but they do not lead to very general 
conclusions. As argued by Mestel (1984) and Tayler (1980) this leaves the 
possibility that special stable configurations might exist with poloidal and 
toroidal fields of similar strength, but no example of such a field is known. 

An indirect, and not completely compelling, argument that stable configurations 
are possible can be made by appealing to the conservation of magnetic helicity 
(Moffat, 1989, private communication). If $\bf A$ is the vector potential of 
$\bf B$ (with a suitable gauge), then it can be shown that the integral of 
$H={\bf B\cdot A}$ over volume is a conserved quantity in ideal MHD (i.e. in the 
absence of magnetic diffusion) (Woltjer 1958, see also Taylor, 1974). If a star 
is born with a field for which $H$ is nonzero, then this field must reach a 
stable equilibrium configuration. Suppose it is not initially in a stable 
equilibrium. By putting in a sufficient damping mechanism such as a viscosity, 
we can make sure that the energy released by the instability is dissipated on 
some finite time scale. The final state must then be a stable equilibrium at a 
finite field strength, since an infinitesimal field can not have a finite 
helicity. This argument is unfortunately not compelling, since it is possible 
that unstable magnetic fields can evolve in such a way as to develop 
singularities (current sheets) within an effectively finite time. Once such 
current sheets develop, reconnection sets in and the helicity is no longer 
conserved. 

While theory fails to give a clear answer to this long standing question (Mestel 
1984), we may appeal to observations to argue that stable configurations do in 
fact exist in stars. The magnetic A stars and the magnetic white dwarfs have 
strong magnetic fields (of the order $10^4$ and $10^7$G, respectively), that do 
not change on time scales of at least decades. Since these stars are also slow 
rotators, the field would change on an Alfv\'en time scale, of the order of a 
year and a day, respectively, if their field configurations were not stable. The 
A stars have convective cores, in which a steady dynamo might possibly  exist to 
produce the observed stable field. The magnetic white dwarfs do not have a 
plausible location for such a dynamo. Very cool white dwarfs may form a 
crystalline lattice in part of the interior, which might be able to anchor a 
strong field. The known magnetic white dwarfs (Schmidt \& Norsworthy 1991, 
Liebert 1995) however, are not all of this type. We can therefore take the 
magnetic white dwarfs as a fairly strong argument for the existence of stable 
field configurations in stars.

Assuming that this is the case, initial field configurations with strength above 
(\ref{B1}) will evolve into stable configurations on the phase mixing time 
scale. The final field strength would depend on the degree to which reconnection 
has taken place during the evolution of the configuration. The star would be 
uniformly rotating, except for special axisymmetric field configurations aligned 
with the rotation axis (which are apparently not realized in the observed 
magnetic A stars and white dwarfs). 

\section{Discussion}

In the above I have reviewed the known processes relevant for magnetic fields in 
differentially rotating stably stratified stellar interiors. This excludes 
dynamo processes such as are thought to occur in convective zones. In initially 
weak fields, (condition \ref{cb1}, about 30G for the Sun) the non-axisymmetric 
components are smoothed out by {\it rotational expulsion}, producing an 
axisymmetric field. Differential rotation is subject to damping by phase mixing. 
 This results in a uniformly rotating star if sufficient time is available, 
compared with time scale on which the internal structure evolves and compared 
with the spin down timescale due to external torques.

The azimuthal field that results from winding-up of an initially weak field is 
subject to instabilities. In the stably stratified environment of a stellar 
interior they are of two types.  There are the Parker (or magnetic buoyancy) and 
Tayler (or stratification-modified pinch-type) instabilities, both driven by the 
magnetic field energy in the toroidal field. They are of particular relevance 
for fields produced by differential rotation, through the winding-up of a weak 
initial field. 

In addition to these instabilities driven by the free energy in the magnetic 
field, there is magnetic shear instability, which feeds on differential 
rotation. It occurs in any field configuration, and occurs already in very weak 
fields. Due to effects of stratification and magnetic diffusion, however, the 
rotation gradient has to be significant for it to occur, especially in slowy 
rotating stars. Poloidal fields configurations have similar instabilities.

It turns out that Tayler instability (pinch-type instability in the presence of 
a stabilizing stratification) is of particular relevance. In an azimuthal field, 
it generically occurs in a region near the pole, in the form of an $m=1$ 
displacement of the field lines along horizontal surfaces. This instability is 
of a local interchange type, and the effects of rapid rotation and magnetic and 
thermal diffusion can be included in the analysis. I find that Acheson's (1978) 
dispersion relation can be applied to this form of instability. The results of 
Sect.\ \ref{horg} and the Appendix show that Tayler instabilities probably are 
more relevant than the better known buoyancy (or Parker-) instabilities. In 
stars with magnetic fields that have been wound up by differential rotation, 
they set in at the lowest field strength.

One can wonder how complete our `catalog' of known instabilities is. As 
discussed above, the situation for nearly azimuthal fields, such as would result 
from differential rotation, is probably quite satisfactory. The energy method 
used by Tayler (1973) gives necessary and sufficient conditions for the 
adiabatic, nonrotating case, and is thus complete for this case. Quite important 
is that this analysis shows that the instabilities are of a {\em local} nature 
(in the $r$ and $\theta$ coordinates). This allows a complete study of the 
effects of rotation, viscosity magnetic and thermal diffusion by Acheson's 
(1978) approach. Barring the possibility of diffusive non-local instabilities 
that don't have an equivalent in the nonrotating adiabiatic case, the stability 
of azimuthal fields can thus be analyzed completely. The same applies 
essentially to purely poloidal fields, where the energy method again shows the 
instabilities to be local. A complete study of the effects of rotation and 
diffusion on these instabilities still has to be done for this case, however. 

Another question is, of course, how the magnetic configuration would evolve 
nonlinearly under these instabilities, in particular how effective they would be 
at transporting angular momentum. It is conceivable that significant progress in 
this question can be made with numerical MHD simulations.

The stability of initially strong fields, such that the differential rotation is 
not strong enough to wind the field into an azimuthal configuration, is as open 
a question as before. The same applies to cases where the field has gone through 
phases of winding-up and phase mixing. In the absence of forces that continue to 
create differential rotation, the phase mixing process eliminates differential 
rotation, and leaves the field in a configuration of unknown stability. If such 
a field is to be stable, it is clear that it can neither be a purely poloidal 
nor a purely toroidal field since these are unstable on short time scales. The 
possibility that a mixed poloidal-toroidal configuration can be stable on long 
time scales can not be excluded (Tayler 1980, Mestel 1984). The magnetic white 
dwarfs are fairly strong observational evidence that such configurations do in 
fact exist.

\begin{acknowledgements}
It is a pleasure to thank Dr. L. Mestel for detailed comments on an earlier 
version of the text, from which the presentation has benefited greatly.
\end{acknowledgements}

\appendix
\section{Appendix: Tayler instability with rotation and diffusion}
We start with the local dispersion relation in Acheson (1978). The goals of the 
analysis are to show that this relation yields the correct results for Tayler 
instability in the known cases, and to extend these to include rotation, 
magnetic and thermal diffusion. 

The perturbations are of the form
\be e^{i(l\vp+m\phi+nz-\om t)}.\ee
The field is axisymmetric and purely azimuthal. Its strength $B$ is taken to be 
a function of cylindrical radius $\vp$ only. If $B$ depends on $z$ as well, 
additional forms of instability are possible (buoyancy or Parker instability). 
This is discussed in Sect.\ \ref{buoy}. At the pole of a uniformly rotating 
star, where gravity is parallel to the rotation axis, and ignoring viscosity, 
the dispersion relation reduces to
\bd
{V_\ra^2\over\vp}\left[2\Om m+\om\left\{2+{l\over n}{\vp\over H}{\om+i\kappa 
s^2\over\om\gamma+i\kappa 
s^2}\right\}\right]\left[\p_hF-{\om\over\om\gamma+i\kappa s^2}\p_hE\right]
\ed
\bd
+ \left[{s^2\over n^2}\left(\om-{\om_\ra^2\over \om+i\eta s^2}\right)+{l\over 
n}{g\over\om\gamma+i\kappa s^2}\p_hE\right]\left[\om(\om+i\eta 
s^2)-\om_\ra^2\right.
\ed
\bd
+\left. {V_\ra^2\over\cs^2}{\om+i\kappa s^2\over\om+i\kappa s^2/\gamma} 
\om^2\right] - \left[2\Om+{mV_\ra^2/\vp^2\over\om+i\eta s^2}(p+1)\right]
\ed
\bd
\times\left[2\Om\left\{\om+i\eta s^2+ {V_\ra^2\over\cs^2}{\om+i\kappa 
s^2\over\om+i\kappa s^2/\gamma}\om\right\} +\right.
\ed
\be
\left.{mV_\ra^2\over\vp^2} \left\{{2\over\vp}+{l\over n}{\vp\over H}{\om+i\kappa 
s^2\over\om\gamma+i\kappa s^2}\right\}\right] =0,
\label{ach0}
\ee
where 
\be 
s^2=n^2+l^2,\qquad \om_\ra=mV_\ra/\vp, 
\ee
\be
 \p_h=\p_\vp-(l/n)\p_z, \qquad c_{\rm s}^2=\gamma P/\rho, 
\ee
\be 
\qquad E=\ln{P\over\rho^\gamma},\qquad F=\ln{B\over\rho\vp}, \qquad p={\rd\ln 
B\over\rd\ln\vp},
\ee
and $H$ the pressure scale height. The magnetic field strengths relevant here 
are weak in the sense $V_\ra\ll c_{\rm s}$, i.e. the magnetic pressure is 
negligible compared to the gas pressure. The terms of order $v_\ra/\cs$ on the 
third and fourth lines can then be neglected, and the gradient of the density in 
the $\vp$-direction can be neglected. Under the same assumption, the expected 
growth rates, of the order $\om_\ra$ or smaller, are small compared with the 
buoyancy frequency $N$, hence instability is possible only for 
$\l/n\sim\om_\ra/N\ll 1$. The terms with factor $l/n$ in the first and fourth 
lines can thus be neglected. The same assumptions also imply that $s/n\approx 
1$. Thus we have
\be \vp\p_hF\approx \rd\ln(B/\vp)\rd\ln\vp=p-1, \ee
\be \p_hE\approx -{l\over n}\p_z\ln(P/\rho^\gamma)=-{l\over n}{N^2\gamma\over 
g}. \ee
The term involving $E$ on the first line can be neglected, but not the term on 
the second line, since it is multiplied by $g$. For $\om\sim\om_\ra$, this term 
is of the same order as the first term on the second line.
Eq. (\ref{ach0}) thus reduces to
\bd
{V_\ra^2\over\vp^2}(2\Om m+2\om)(p-1)+ 
\ed
\bd
\left[\om-{\om_\ra^2\over\om+i\eta s^2} -{l^2\over n^2}{N^2\over\om+i\kappa 
s^2/\gamma}\right]\left[\om(\om+i\eta s^2)-\om_\ra^2\right] -
\ed
\be \left[2\Om+{mV_\ra^2/\vp^2\over\om+i\eta s^2}(p+1)\right] 
\left[2\Om(\om+i\eta s^2)+2{mV_\ra^2\over\vp^2}\right] =0.
\label{ach1}
\ee

To see that the stability conditions implied by this relation are the same, in 
the adiabatic case as, those found by Tayler (1973), Goossens et al. (1981), and 
Pitts \& Tayler (1986), set $\eta=\kappa=0$, so that
\bd
{V_\ra^2\over\vp^2}(2\Om\om m+2\om^2)(p-1)+ \left[\om^2-\om_\ra^2 -{l^2\over 
n^2}N^2\right](\om^2-\om_\ra^2) 
\ed
\be 
- \left[2\Om\om+{mV_\ra^2\over \vp^2}(p+1)\right] 
\left[2\Om\om+2{mV_\ra^2\over\vp^2}\right] =0.
\label{ach2}
\ee
Consider first the case $\Om=0$. Then (\ref{ach2}) is a quadratic equation in 
$\om^2$. Instability exists if there are wavenumbers for which $\om^2<0$. The 
sufficient and necessary condition for instability is that the constant term in 
the quadratic is positive. The cases $m=0$ and $m\ne 0$ have to be considered 
separately. For both cases, the range of instability is maximized in the limit 
in which the vertical wavenumber $n\rightarrow\infty$, and the instability 
conditions are
\be p>1\qquad (m=0),\qquad p>{m^2\over 2}-1\qquad (m\ne 0). \label{adi0}\ee
These are the conditions found by Tayler (1973) and Goossens et al. (1981), if 
we take into account that their results are valid on the entire sphere, whereas 
we have considered only the situation near the pole.

For an azimuthal field resulting from the winding up of a a radial field 
component by differential rotation, $B\sim\varpi$ near the pole, so that the 
$m=0$ mode is
marginally stable. The higher $m$'s require a steeper field gradient than $m=1$, 
hence $m=1$ is the most unstable mode, at least in some region around the pole. 
From now on I ignore the $m=0$ mode. 

The case of arbitrary rotation rate is a bit complicated, so I consider only the 
limiting cases $\Om=0$ and $\Om/\om_\ra\rightarrow\infty$. For $\Om=0$, the 
dispersion relation can be made real by the substitution
\be \om=i\sigma.\ee
Let $K\equiv\kappa s^2/\gamma$ and $H\equiv\eta s^2$. Multiplying (\ref{ach1}) 
by $m^2(\si+K)(\si+H)$, one gets a fifth degree polynomial equation in $\si$:
\bd
-\om_\ra^2(p-1)\si(\si+H)(\si+K)+{m^2\over 2}[\si(\si+H)(\si+K)+
\ed
\bd
\om_\ra^2(\si+K) +{l^2\over n^2}N^2(\si+H)][\si(\si+H)+\om_\ra^2]
\ed
\be
-\om_\ra^4(\si+K)(p+1)=0.~(\Om=0)\label{om0a}
\ee
The system can in principle have both monotonic and oscillatory (overstable) 
instabilities, and Eq. (\ref{om0a}) would have to be checked for both 
possibilities. For systems with real coefficients like (\ref{om0a}), experience 
from double diffusive systems shows that overstability results if the 
destabilizing agent diffuses faster than the stabilizing agent, while monotonic 
instability results if the destabilizing agent has the lower diffusivity. In the 
present case the destabilizing agent is the magnetic field, which has a lower 
diffusivity than the stabilizing thermal stratification, so we expect monotonic 
instability. Marginal stability then corresponds to $\sigma=0$, and the 
condition for instability is that the constant term in the polynomial be 
negative. This yields
\be 
p>{m^2\over 2}-1+{l^2\over n^2}{\gamma N^2\over\om_\ra^2}{\eta\over\kappa}. 
\label{om0}
\ee
For a smooth field gradient, $p\sim O(1)$, instability is possible only if the 
last term does not exceed order unity:
\be  n^2>l^2{N^2\over\om_\ra^2}{\eta\over\kappa}.\ee
This can be achieved, for arbitrarily low field strength, by taking the radial 
wavenumber $n$ sufficiently large. There is no critical field strength for 
instability. Magnetic diffusion does have an effect on the growth rates, 
however. Closer inspection of Eq. (\ref{om0a}) shows that the maximum growth 
rate as a function of $n$ is of the order of the adiabatic rate $\om_\ra$ only 
if a critical field strength $\om_{\rm Ac}$ is exceeded, given by:
\be \om_{\rm Ac}^3 \approx{\eta^2\over r^2\kappa}N^2,\label{om0b}\ee
where we have taken $l\sim 1/r$ as the lowest possible horizontal wavenumber. If 
this satisfied, the radial wavenumber at maximum growth rate is of the order 
$n\sim (\om_\ra/\eta)^{1/2}$. If the field strength is less than given by 
(\ref{om0b}), the maximum growth rate is reduced:
\be 
\si\sim\om_\ra\quad(\om_\ra\gg\om_{\rm Ac}),\qquad\si\sim\om_\ra^2 
{r^2\over\eta} \quad (\om_\ra\ll\om_{\rm Ac}).
\ee

Next consider the opposite limiting case, $\Om\gg\om_\ra$.  In this limit, one 
finds that the frequency scales as $\om\sim\om_\ra^2/\Om$. Writing
\be
\om=\al\om_\ra^2/\Om, \qquad h={\eta\Om\over\om_\ra^2}, \qquad
k={\kappa\Om\over\gamma\om_\ra^2},
\ee
and neglecting higher order terms in $\om_\ra/\Om$,
relation (\ref{ach1}) reduces to
\bd
m(p-1)(\al+in^2h)(\al+in^2k)+
\ed
\bd
{m^2\over 2}[\al+in^2k+{l^2N^2\over n^2\om_\ra^2}(\al+in^2h)]-
\ed
\be
[2m(\al+in^2h)+p+1](\al+in^2k)[m(\al+in^2h)+1]=0.
\label{arot}
\ee

Consider first the adiabatic case, $h=k=0$. The dispersion relation (\ref{arot}) 
 is then quadratic in $\al$ and one finds that the necessary and sufficient 
condition for instability is
\be p>1+{m^2\o 2}. \qquad(m\ne 0,~\Om\gg\om_\ra).\ee
This condition is significantly more restrictive than nonrotating condition 
(\ref{adi0}). For a field $B\sim\vp$, such as would result from differential
rotation near the pole, the condition predicts stability. Since the fields we 
envisage are just of this type, one would conclude stabilily, at least in the 
interesting region near the pole that critical for driving the instability. This 
was noted by Pitts \& Tayler (1986), who also found that a sufficiently large 
gradient $p$ would again be unstable in the rapidly rotating case. It turns out, 
as shown next, that the instability condition for the rapidly rotating case is 
relaxed again when the effects of thermal and magnetic diffusion are taken into 
account.

Returning to Eq. (\ref{arot}) one easily verifies that there is no direct 
instability ($\al$ imaginary), since the coefficients are complex due to the 
combined effects of rotation and diffusion.  Thus we have to check the stability 
of oscillatory modes. The stability boundary for marginally stable oscillations 
is found by requiring the dispersion relation to have a solution with $\al$ 
real. The case with both diffusivities present is rather complicated, so I 
specialize further to the cases
$\kappa=0$ and $\kappa/\eta\rightarrow\infty$. The first limit is appropriate 
for cases where the stratification $N^2$ is due to a composition gradient 
instead of the thermal gradient, because the ions making up such a gradient 
diffuse only slowly. The latter case applies when composition gradients can be 
ignored.

For the case $k=0$ the real and imaginary parts of (\ref{arot}) yield
\bea
Re:&\quad -2(m\al+1)^2+2+{m^2\over 2}-(p+1)+\nonumber\cr
  &{m^2\over 2}{l^2N^2\over n^2\om_\ra^2}+2m^2n^4h^2=0,\cr
Im:&\quad -2m^2\al^2-2m\al+{m^2\over 4}{l^2N^2\over n^2\om_\ra^2}=0,\cr
&(\kappa=0,~\Om\gg\om_\ra)\label{rim}
\eea
Eliminating $\al$ between the real and imaginary parts yields
\be 
\half[(1+{m^2\over 2}{l^2N^2\over n^2\om_\ra^2})^{1/2}-1]^2+{m^2\over 
2}-(p+1)+2m^2n^4h^2=0.\label{k0}
\ee
The terms involving the wavenumber $n$ are both positive. The last one increases 
monotonically with wave number, the first with the inverse of the wavenumber. 
Instability, if it exists, is therefore restricted to a finite range in 
wavenumbers. At the high wavenumber end the instability is cut short by magnetic 
diffusion (last term), at low wavenumbers by the stable stratification (first 
term). Instability exists if there are wavenumbers for which Eq. (\ref{k0}) has 
solutions. Necessary and sufficient for this to be the case is that the sum of 
the first and last terms be less than $p+1-m^2/2$. This yields rather 
complicated expressions. Instead of the exact conditions, a sufficient condition 
for instability which is also sufficiently accurate as a necessary condition is 
found by noting that $[(1+x^2)^{1/2}-1]\le x^2$, hence instability is guaranteed 
if there are $n$'s for which both
\be
{m^2\o 4}{l^2N^2\over n^2\om_\ra^2}<\half(p+1-{m^2\o 2})
\ee
and
\be 
2m^2n^4h^2<\half(p+1-{m^2\o 2}).
\ee
This is possible only if the adiabatic instability condition $p+1-m^2/2>0$ is 
satisfied, hence magnetic diffusion restricts the range of instability (as 
opposed to other double diffusive systems, where diffusion can be 
destabilizing).
Both conditions can be satisfied if both $p>{m^2\o 2}-1$ and
\be \om_\ra^4>{\vert m\vert^3\over a^{3/2}}l^2N^2\eta\Om,\qquad 
(\kappa=0,~\Om\gg\om_\ra)\ee
where 
\be a=p+1-m^2/2.\label{aa}\ee
With $m=1$, $l\sim 1/r$, and $a\sim 1$, this is equivalent to the heuristic 
result of Sect.\ \ref{tay} (eq {\ref{dif1}).

In the opposite limit $\kappa\gg\eta$, the real and imaginary parts of  
(\ref{arot}) yield
\bea
&2-2(m\al+1)^2+f^2 +{m^2\over 2}-(p+1)+g^2=0,\cr
&m\al=-2g^2/(f^2+2g^2),
\eea
where
\be
f^2={m^2\over 2}{l^2N^2\over n^2\om_\ra^2} {\eta\over\kappa},\qquad 
g^2=2m^2n^4h^2.
\ee
The second of these implies that
\be 2-2(m\al+1)^2>0.\ee
Necessary for instability is again that $a=p+1-m^2/2>0$. Supposing that this is 
satisfied, a sufficient for instability is that there exist wavenumbers for 
which 
\be 
2-2(m\al+1)^2<{1\o 3}a,\quad{\rm and}~~ f^2<{1\o 3}a,\quad{\rm and}~~ g^2<{1\o 
3}a.
\ee 
This is the case if both $p>{m^2\o 2}-1$ and
\be 
\om_\ra^4\gapprox{\vert m\vert^3\over a^{3/2}} l^2N^2{\eta\over\kappa} 
\eta\Om,\qquad (\eta\ll\kappa,~\Om\gg\om_\ra).\label{adif2}
\ee
This is a slightly imprecise condition, since I have not determined the exact 
numerical factor in front of the RHS. With $m=1$, $l\sim 1/r$, and $a\sim 1$, 
this condition is equivalent to the heuristic result of Sect.\ \ref{tay} (eq 
\ref{dif2}).

\end{document}